\pdfoutput=1
\PassOptionsToPackage{pdfpagelabels=false}{hyperref} 
\documentclass{sigchi}
\usepackage[usenames,dvipsnames,svgnames,table]{xcolor}
\usepackage{balance}
\usepackage{graphics}
\usepackage{txfonts}
\usepackage{times}
\usepackage[pdftex]{hyperref}
\usepackage{url}
\usepackage{color}
\usepackage{textcomp}
\usepackage{booktabs}
\usepackage{ccicons}
\usepackage{siunitx}

\usepackage[inline]{enumitem}
\usepackage{quoting}
\usepackage{multirow}
\usepackage{booktabs}
\usepackage{dcolumn}
\newcommand{\ra}[1]{\renewcommand{\arraystretch}{#1}}
\usepackage{subcaption}
\newcolumntype{.}{D{.}{.}{-1}}
\usepackage[all]{nowidow}

\makeatletter
\def\url@leostyle{%
  \@ifundefined{selectfont}{\def\UrlFont{\sf}}{\def\UrlFont{\small\bf\ttfamily}}}
\makeatother
\def\pprw{8.5in}
\def\pprh{11in}

\definecolor{linkColor}{RGB}{6,125,233}
\hypersetup{%
  pdftitle={SIGCHI Conference Proceedings Format},
  pdfauthor={LaTeX},
  pdfkeywords={SIGCHI, proceedings, archival format},
  bookmarksnumbered,
  pdfstartview={FitH},
  colorlinks,
  citecolor=black,
  filecolor=black,
  linkcolor=black,
  urlcolor=linkColor,
  breaklinks=true,
}

\usepackage{xspace}
\newcommand\test{quiz\xspace}
\newcommand\Test{Quiz\xspace}

\newcommand{\xhdr}[1]{\vspace{1mm}\noindent{{\bf #1.}}}

\graphicspath{{./figures/}}

\toappear{\scriptsize Permission to make digital or hard copies of all or part of this work for personal or classroom use is granted without fee provided that copies are not made or distributed for profit or commercial advantage and that copies bear this notice and the full citation on the first page. Copyrights for components of this work owned by others than ACM must be honored. Abstracting with credit is permitted. To copy otherwise, or republish, to post on servers or to redistribute to lists, requires prior specific permission and/or a fee. Request permissions from permissions@acm.org. \\
{\emph{CSCW 2017, February 25--March 1, 2017, Portland, OR, USA.} } \\
Copyright is held by the owner/author(s). Publication rights licensed to ACM. \\
ACM ISBN 978-1-4503-4189-9/16/10\ ...\$15.00.\\
http://dx.doi.org/10.1145/2998181.2998213}

\begin{document}

\title{Anyone Can Become a Troll:\\Causes of Trolling Behavior in Online Discussions}

\numberofauthors{1}
\author{Justin Cheng$^1$, Michael Bernstein$^1$, Cristian Danescu-Niculescu-Mizil$^2$, Jure Leskovec$^1$ \\
\affaddr{$^1$Stanford University, $^2$Cornell University}\\
\email{\{jcccf, msb, jure\}@cs.stanford.edu, cristian@cs.cornell.edu}
}

\maketitle

\begin{abstract}
In online communities, antisocial behavior such as trolling disrupts constructive discussion.
While prior work suggests that trolling behavior is confined to a vocal and antisocial minority, we demonstrate that ordinary people can engage in such behavior as well.
We propose two primary trigger mechanisms: the individual's mood, and the surrounding context of a discussion (e.g., exposure to prior trolling behavior).
Through an experiment simulating an online discussion, we find that both negative mood and seeing troll posts by others significantly increases the probability of a user trolling, and together double this probability.
To support and extend these results, we study how these same mechanisms play out in the wild via a data-driven, longitudinal analysis of a large online news discussion community.
This analysis reveals temporal mood effects, and explores long range patterns of repeated exposure to trolling.
A predictive model of trolling behavior
shows
that mood and discussion context together can
explain trolling behavior better than an individual's history of trolling.
These results combine to suggest that ordinary people can, under the right circumstances, behave like trolls.
\end{abstract}



\section{Introduction}
\label{sec:intro}
As online discussions become increasingly part of our daily interactions \cite{diakopoulos2011towards}, antisocial behavior such as trolling \cite{hardaker2010trolling,kayany1998contexts}, harassment, and bullying \cite{slonje2013nature} is a growing concern.
Not only does antisocial behavior result in significant emotional distress \cite{akbulut2010cyberbullying,li2005cyber,raskauskas2007involvement}, but it can also lead to offline harassment and threats of violence \cite{wiener1998negligent}.
Further, such behavior comprises a substantial fraction of user activity on many web sites \cite{cheng2015antisocial,diakopoulos2011towards,gardiner2016trolls} -- 40\% of internet users were victims of online harassment \cite{pew2014online}; on CNN.com, over one in five comments are removed by moderators for violating community guidelines.
What causes this prevalence of antisocial behavior online?

In this paper, we focus on the causes of {\em trolling behavior} in discussion communities, defined in the literature as behavior that falls outside acceptable bounds defined by those communities~\cite{binns2012don,cnnguidelines,hardaker2010trolling}.
Prior work argues that trolls are born and not made: those engaging in trolling behavior have unique personality traits \cite{buckels2014trolls} and motivations \cite{baker2001moral,herring2002searching,shachaf2010beyond}.
However, other research suggests that people can be influenced by their environment to act aggressively \cite{cialdini2004social,jones1978air}.
As such, is trolling caused by particularly antisocial individuals or by ordinary people?
Is trolling behavior innate, or is it situational?
Likewise, what are the conditions that affect a person's likelihood of engaging in such behavior?
And if people can be influenced to troll, can trolling spread from person to person in a community?
By understanding what causes trolling and how it spreads in communities, we can design more robust social systems that can guard against such undesirable behavior.

This paper reports a field experiment and observational analysis of trolling behavior in a popular news discussion community.
The former allows us to tease apart the causal mechanisms that affect a user's likelihood of engaging in such behavior.
The latter lets us replicate and
explore finer grained aspects of
 these mechanisms as they occur in the wild.
Specifically, we focus on two possible causes of trolling behavior: a user's mood, and the surrounding discussion context (e.g., seeing others' troll posts before posting).

\xhdr{Online experiment}
We studied the effects of participants' prior mood and the context of a discussion on their likelihood to leave troll-like comments.
Negative mood increased the probability of a user subsequently trolling in an online news comment section, as did the presence of prior troll posts written by other users.
These factors combined to double participants' baseline rates of engaging in trolling behavior.

\xhdr{Large-scale data analysis}
We augment these results with an analysis of over 16 million posts on \emph{CNN.com}, a large online news site where users can discuss published news articles.
One out of four posts flagged for abuse are authored by users with no prior record of such posts, suggesting that many undesirable posts can be attributed to ordinary users.
Supporting our experimental findings, we show that a user's propensity to troll rises and falls in parallel with known population-level mood shifts throughout the day \cite{golder2011diurnal}, and exhibits cross-discussion persistence and temporal decay patterns, suggesting that negative mood from bad events linger \cite{jones1978air,kendall1975timeout}.
Our data analysis also recovers the effect of exposure to prior troll posts in the discussion, and further reveals how the strength of this effect depends on the volume and ordering of these posts.

Drawing on this evidence, we develop a logistic regression model that accurately (AUC=0.78) predicts whether an individual will troll in a given post.
This model also lets us evaluate the relative importance of mood and discussion context, and contrast it with prior literature's assumption of trolling being innate.
The model reinforces our experimental findings -- rather than trolling behavior being mostly intrinsic, such behavior can be mainly explained by the discussion's context (i.e., if prior posts in the discussion were flagged), as well as the user's mood as revealed through their recent posting history (i.e., if their last posts in other discussions were flagged).

Thus, not only can negative mood and the surrounding discussion context prompt ordinary users to engage in trolling behavior, but such behavior can also spread from person to person in discussions and persist across them to spread further in the community.
Our findings suggest that trolling, like laughter, can be contagious, and that ordinary people, given the right conditions, can act like trolls.
In summary, we:
\begin{itemize}[noitemsep]
\item present an experiment that shows that both negative mood and discussion context increases the likelihood of trolling,
\item validate these findings with a large-scale analysis of a large online discussion community, and
\item use these insights to develop a predictive model that suggests that trolling may be more situational than innate.
\end{itemize}

\section{Background}
\label{sec:related}
To begin, we review literature on antisocial behavior (e.g., aggression and trolling) and influence (e.g., contagion and cascading behavior), and identify open questions about how trolling spreads in a community.

\subsection{Antisocial behavior in online discussions}
Antisocial behavior online can be seen as an extension of similar behavior offline, and includes acts of aggression, harassment, and bullying \cite{akbulut2010cyberbullying,kayany1998contexts}.
Online antisocial behavior increases anger and sadness \cite{li2005cyber}, and threatens social and emotional development in adolescents \cite{raskauskas2007involvement}.
In fact, the pain of verbal or social aggression may also linger longer than that of physical aggression \cite{chen2008hurt}.

Antisocial behavior can be commonly observed in online public discussions, whether on news websites or on social media.
Methods of combating such behavior include comment ranking \cite{hsu2009ranking}, moderation \cite{lampe2004slash,park2016supporting}, early troll identification \cite{chancellor2016post,cheng2015antisocial}, and interface redesigns that encourage civility \cite{kriplean2012supporting,kriplean2012you}.
Several sites have even resorted to completely disabling comments \cite{finley2015brief}.
Nonetheless, on the majority of popular web sites which continue to allow discussions, antisocial behavior continues to be prevalent \cite{cheng2015antisocial,diakopoulos2011towards,gardiner2016trolls}.
In particular, a rich vein of work has focused on understanding trolling on these discussion platforms \cite{donath1999identity, hardaker2010trolling}, for example discussing the possible causes of malicious comments \cite{lee2015people}.

A troll has been defined in multiple ways in previous literature -- as a person who initially pretends to be a legitimate participant but later attempts to disrupt the community \cite{donath1999identity}, as someone who ``intentionally disrupts online communities''~\cite{schwartz2008trolls}, or ``takes pleasure in upsetting others'' \cite{kirman2012exploring}, or more broadly as a person engaging in ``negatively marked online behavior'' \cite{hardaker2010trolling} or that ``makes trouble'' for a discussion forums' stakeholders \cite{binns2012don}.
In this paper, similar to the latter studies, we adopt a definition of trolling that includes flaming, griefing, swearing, or personal attacks, including behavior outside the acceptable bounds defined by several community guidelines for discussion forums \cite{cnnguidelines,discourseguidelines,guardianguidelines}.\footnote{In contrast to cyberbullying, defined as behavior that is repeated, intended to harm, and targeted at specific individuals \cite{slonje2013nature}, this definition of trolling encompasses a broader set of behaviors that may be one-off, unintentional, or untargeted.}
In our experiment, we code posts manually for trolling behavior.
In our longitudinal data analysis, we use posts that were flagged for unacceptable behavior as a proxy for trolling behavior.

Who engages in trolling behavior?
One popular recurring narrative in the media suggests that trolling behavior comes from {\em trolls}: a small number of particularly sociopathic individuals \cite{rensin2014trolls,schwartz2008trolls}.
Several studies on trolling have focused on a small number of individuals \cite{baker2001moral,binns2012don,herring2002searching,shachaf2010beyond}; other work shows that there may be predisposing personality (e.g., sadism \cite{buckels2014trolls}) and biological traits (e.g., low baseline arousal \cite{raine2002annotation}) to aggression and trolling. That is, trolls are born, not made.

Even so, the prevalence of antisocial behavior online suggests that these trolls, being relatively uncommon, are not responsible for all instances of trolling.
Could ordinary individuals also engage in trolling behavior, even if temporarily?
People are less inhibited in their online interactions \cite{suler2004online}.
The relative anonymity afforded by many platforms also deindividualizes and reduces accountability \cite{zimbardo1969human}, decreasing comment quality \cite{kilner2005anonymity}.
This disinhibition effect suggests that people, in online settings, can be more easily influenced to act antisocially.
Thus, rather than assume that only trolls engage in trolling behavior, we ask:
\textit{RQ: Can situational factors trigger trolling behavior?}

\subsection{Causes of antisocial behavior}
Previous work has suggested several motivations for engaging in antisocial behavior: out of boredom \cite{varjas2010high}, for fun \cite{shachaf2010beyond}, or to vent \cite{lee2015people}.
Still, this work has been largely qualitative and non-causal, and whether these motivations apply to the general population remains largely unknown.
Out of this broad literature, we identify two possible trigger mechanisms of trolling -- mood and discussion context -- and try to establish their effects using both a controlled experiment and a large-scale longitudinal analysis.

\xhdr{Mood} Bad moods may play a role in how a person later acts.
Negative mood correlates with reduced satisfaction with life~\cite{schwarz1983mood}, impairs self-regulation \cite{leith1996bad}, and leads to less favorable impressions of others \cite{forgas1987mood}.
Similarly, exposure to unrelated aversive events (e.g., higher temperatures \cite{rotton1985air} or secondhand smoke \cite{jones1978air}) increases aggression towards others.
An interview study found that people thought that malicious comments by others resulted from ``anger and feelings of inferiority'' \cite{lee2015people}.

Nonetheless, negative moods elicit greater attention to detail and higher logical consistency \cite{schwarz1991happy}, which suggests that people in a bad mood may provide more thoughtful commentary.
Prior work is also mixed on how affect influences prejudice and stereotyping.
Both positive \cite{bodenhausen1994happiness} and negative affect \cite{greenberg1992terror} can increase stereotyping, and thus trigger trolling \cite{herring2002searching}.
Still, we expect the negative effects of negative mood in social contexts to outweigh these other factors.

Circumstances
that influence mood may also modify the rate of trolling.
For instance, mood changes with the time of day or day of week \cite{golder2011diurnal}.
As negative mood rises at the start of the week, and late at night, trolling may vary similarly.
``Time-outs'' or allowing for a period of calming down \cite{kendall1975timeout} can also reduce aggression -- users who wait longer to post after a bout of trolling may also be less susceptible to future trolling.
Thus, we may be able to observe how mood affects trolling, directly through experimentation, and indirectly through observing factors that influence mood:

\textit{H1: Negative mood increases a user's likelihood of trolling.}

\xhdr{Discussion context}
A discussion's context may also affect what people contribute.
The discussion starter influences the direction of the rest of the discussion \cite{hara2000content}. Qualitative analyses suggest that people think online commenters follow suit in posting positive (or negative) comments \cite{lee2015people}.
More generally, standards of behavior (i.e., social norms) are inferred from the immediate environment \cite{chang2016engineering,cialdini2004social,milgram1978obedience}.
Closer to our work is an experiment that demonstrated that less thoughtful posts led to less thoughtful responses \cite{sukumaran2011normative}.
We extend this work by studying amplified states of antisocial behavior (i.e., trolling) in both experimental and observational settings.

On the other hand, users may not necessarily react to trolling with more trolling.
An experiment that manipulated the initial votes an article received found that initial downvotes tended to be corrected by the community \cite{muchnik2013social}.
Some users respond to trolling with sympathy or understanding \cite{baker2001moral}, or apologies or joking \cite{lee2005behavioral}.
Still, such responses are rarer \cite{baker2001moral}.

Another aspect of a discussion's context is the subject of discussion.
In the case of discussions on news sites, the topic of an article can affect the amount of abusive comments posted \cite{gardiner2016trolls}.
Overall, we expect that previous troll posts, regardless of who wrote them, are likely to result in more subsequent trolling, and that the topic of discussion also plays a role:

\textit{H2: The discussion context (e.g., prior troll posts by other users)
affects
a user's likelihood of trolling.}

\subsection{Influence and antisocial behavior}
That people can be influenced by environmental factors suggests that trolling could be contagious -- a single user's outburst might lead to multiple users participating in a flame war.
Prior work on social influence \cite{banerjee1992simple} has demonstrated multiple examples of herding behavior, or that people are likely to take similar actions to previous others \cite{cialdini1990focus,milgram1969note,zimbardo1969human}.
Similarly, emotions and behavior can be transferred from person to person \cite{barsade2002ripple,centola2010spread,gino2009contagion,kramer2014experimental,wheeler1966toward}.
More relevant is work showing that getting downvoted leads people to downvote others more and post content that gets further downvoted in the future \cite{cheng2014community}.

These studies generally point toward a ``Broken Windows'' hypothesis, which postulates that untended behavior can lead to the breakdown of a community \cite{wilson1982broken}.
As an unfixed broken window may create a perception of unruliness, comments made in poor taste may invite worse comments.
If antisocial behavior becomes
the norm,
this can lead a community to further perpetuate it despite its undesirability \cite{willer2009false}.

Further evidence for the impact of antisocial behavior stems from research on negativity bias -- that negative traits or events tend to dominate positive ones.
Negative entities are more contagious than positive ones \cite{rozin2001negativity}, and bad impressions are quicker to form and more resistant to disconfirmation \cite{baumeister2001bad}.
Thus, we expect antisocial behavior is particularly likely to be influential, and likely to persist.
Altogether, we hypothesize:

\textit{H3: Trolling behavior can spread from user to user.}

We test H1 and H2 using a controlled experiment, then verify and extend our results with an analysis of discussions on CNN.com.
We test H3 by studying the evolution of discussions on CNN.com, finally developing an overall model for how trolling might spread from person to person.

\section{Experiment: Mood and Discussion Context}
\label{sec:experiment}
\begin{figure*}[tb]
    \centering
    \begin{tabular}{@{}cc@{}}
    \hspace{-0.1cm}
    \includegraphics[width=0.48\textwidth]{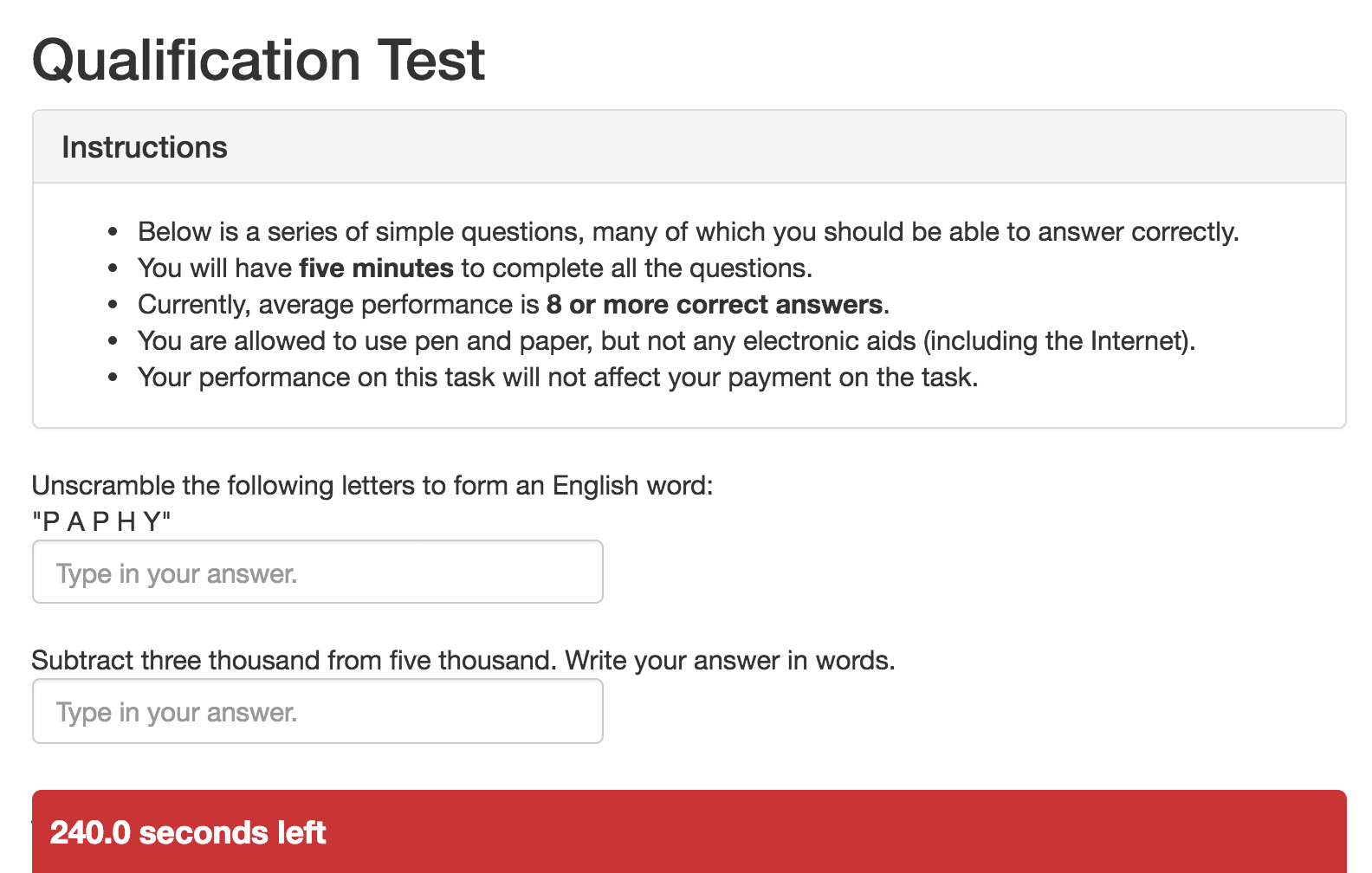}
    \hspace{-0.1cm}
    &
    \hspace{-0.1cm}
    \includegraphics[width=0.48\textwidth]{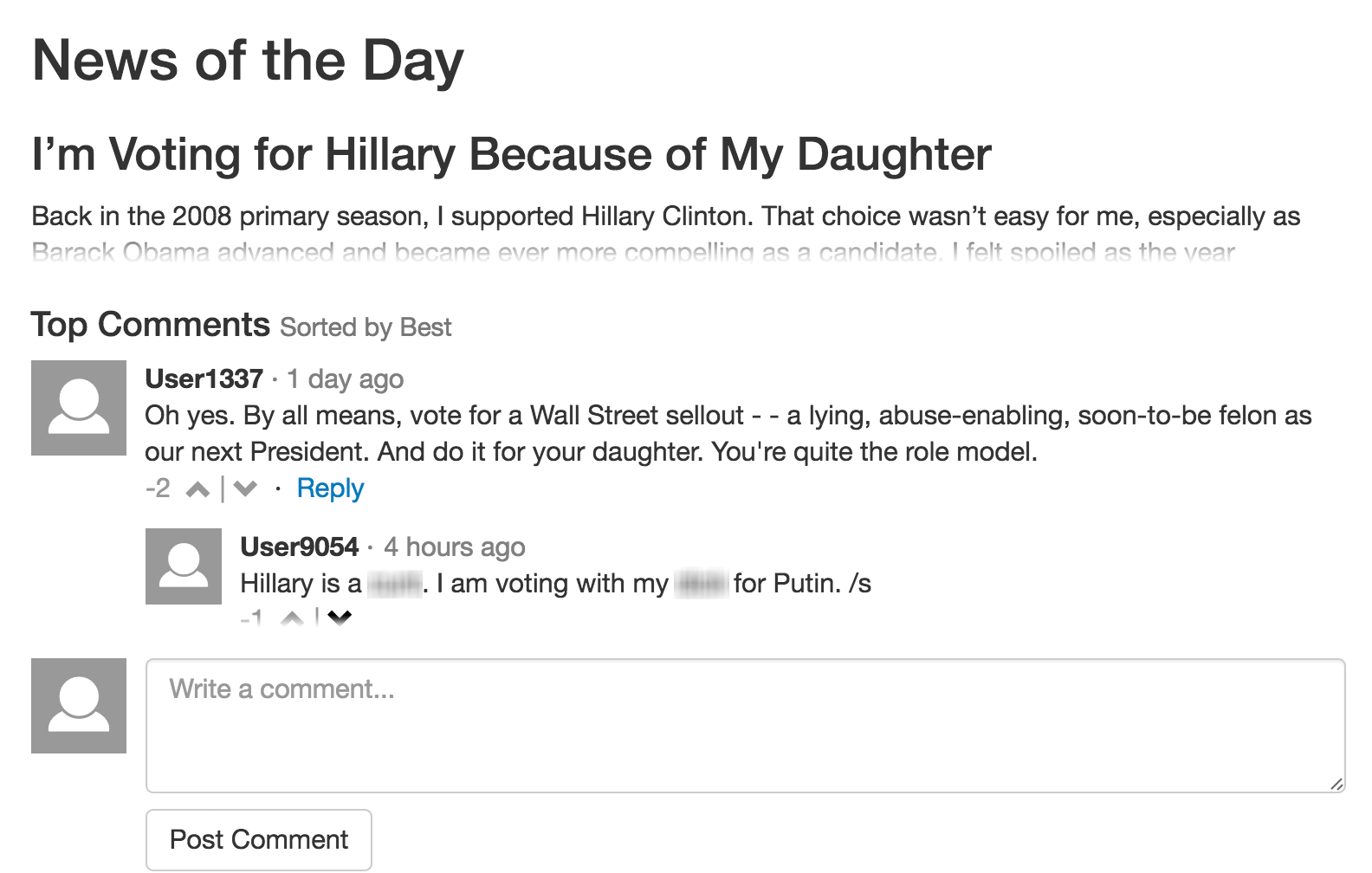}\\
    \small{(a)} & \small{(b)}
    \end{tabular}
    \caption{
    To understand how a person's mood and discussion's context (i.e., prior troll posts) affected the quality of a discussion, we conducted an experiment that varied (a) how difficult a \test, given prior to participation in the discussion, was, as well as (b)~ whether the initial posts in a discussion were troll posts or not.
    }
    \label{fig:interface}
\end{figure*}

To establish the effects of mood and discussion context, we deployed an experiment designed to replicate a typical online discussion of a news article.

Specifically, we measured the effect of mood and discussion context on the quality of the resulting discussion across two factors:
\begin{enumerate*}[label={\alph*)}]
\item \textsc{PosMood} or \textsc{NegMood}: participants were either exposed to an unrelated positive or negative prior stimulus (which in turn affected their prevailing mood), and
\item \textsc{PosContext} or \textsc{NegContext}: the initial posts in the discussion thread were either benign (or not troll-like), or troll-like.
\end{enumerate*}
Thus, this was a two-by-two between-subjects design, with participants assigned in a round robin to each of the four conditions.

We evaluated discussion quality using two measures:
\begin{enumerate*}[label={\alph*)}]
\item trolling behavior, or whether participants wrote more troll-like posts, and
\item affect, or how positive or negative the resulting discussion was, as measured using sentiment analysis.
\end{enumerate*}

If negative mood (\textsc{NegMood}) or troll posts (\textsc{NegContext}) affects the probability of trolling, we would expect these conditions to reduce discussion quality.

\subsection{Experimental Setup}
The experiment consisted of two main parts -- a \test, followed by a discussion -- and was conducted on Amazon Mechanical Turk (AMT).
Past work has also recruited workers to participate in experiments with online discussions \cite{mcinnis2016one}.
Participants were restricted to residing in the US, only allowed to complete the experiment once, and compensated \$2.00, for an hourly rate of \$8.00.
To avoid demand characteristics, participants were not told of the experiment's purpose prior, and were only instructed to complete a
\test, and then participate in an online discussion.
After the experiment, participants were debriefed and told of its purpose (i.e., to measure the impact of mood and trolling in discussions).
The experimental protocol was reviewed and conducted under IRB Protocol \#32738.

\xhdr{\Test (\textsc{PosMood} or \textsc{NegMood})}
The goal of the \test was to see if participants' mood prior to participating in a discussion had an effect on subsequent trolling.
Research on mood commonly involves giving people negative feedback on tasks that they perform in laboratory experiments regardless of their actual performance \cite{knobloch2006mood,wyland2007bad,gonzalez2012investigating}.
Adapting this to the context of AMT, where workers care about their performance on tasks and qualifications (which are necessary to perform many higher-paying tasks), participants were instructed to complete an experimental test qualification that was being considered for future use on AMT.
They were told that their performance on the \test would have no bearing on their payment at the end of the experiment.

The \test consisted of 15 open-ended questions, and included logic, math, and word problems (e.g., word scrambles) (Figure \ref{fig:interface}a).
In both conditions, participants were given five minutes to complete the \test, after which all input fields were disabled and participants forced to move on.
In both the \textsc{PosMood} and \textsc{NegMood} conditions, the composition and order of the types of questions remained the same.
However, the \textsc{NegMood} condition was made up of questions that were substantially harder to answer within the time limit: for example, unscramble ``DEANYON'' (\textsc{NegMood}) vs. ``PAPHY'' (\textsc{PosMood}).
At the end of the \test, participants' answers were automatically scored, and their final score displayed to them.
They were told whether they performed better, at, or worse than the ``average'', which was fixed at eight correct questions.
Thus, participants were expected to perform well in the \textsc{PosMood} condition and receive positive feedback, and expected to perform poorly in the \textsc{NegMood} condition and receive negative feedback, being told that they were performing poorly, both absolutely and relatively to other users.
While users in the \textsc{PosMood} condition can still perform poorly, and users in the \textsc{NegMood} condition perform well, this only reduces the differences later observed.

To measure participants' mood following the \test, and acting as a manipulation check, participants then completed 65 Likert-scale questions on how they were feeling based on the Profile of Mood States (POMS) questionnaire \cite{mcnair1971manual}, which quantifies mood on six axes such as anger and fatigue.

\begin{table}[tb]
\small
\centering
\ra{1.3}
\begin{tabular*}{\columnwidth}{@{\extracolsep{\fill} }lll|ll}
\toprule
& \multicolumn{2}{c}{Proportion of Troll Posts} & \multicolumn{2}{c}{Negative Affect (LIWC)} \\
& \textsc{PosMood} & \textsc{NegMood} & \textsc{PosMood} & \textsc{NegMood} \\
\textsc{PosContext} & 35\% & 49\% & 1.1\% & 1.4\% \\
\textsc{NegContext} & 47\% & \textbf{68\%} & 2.3\% & \textbf{2.9\%} \\
\bottomrule
\end{tabular*}
\caption{The proportion of user-written posts that were labeled as trolling
(and proportion of words with negative affect)
 was lowest in the (\textsc{PosMood}, \textsc{PosContext}) condition, and highest, and almost double, in the (\textsc{NegMood}, \textsc{NegContext}) condition
 (highlighted in bold).
 }
\label{tab:exp_values}
\end{table}

\xhdr{Discussion (\textsc{PosContext} or \textsc{NegContext})}
Participants were then instructed to take part in an online discussion, and told that we were testing a comment ranking algorithm.
Here, we showed participants an interface similar to what they might see on a news site --- a short article, followed by a comments section.
Users could leave comments, reply to others' comments, or upvote and downvote comments (Figure \ref{fig:interface}b).
Participants were required to leave at least one comment, and told that their comments may be seen by other participants.
Each participant was randomly assigned a username (e.g., User1234) when they commented.
In this experiment, we showed participants an abridged version of an article arguing that women should vote for Hillary Clinton instead of Bernie Sanders in the Democratic primaries leading up to the 2016 US presidential election \cite{june2016hillary}.
In the \textsc{NegContext} condition, the first three comments were troll posts, e.g.,:

\begin{quoting}
\emph{Oh yes. By all means, vote for a Wall Street sellout -- a lying, abuse-enabling, soon-to-be felon as our next President. And do it for your daughter. You're quite the role model.}
\end{quoting}

In the \textsc{PosContext}, they were more innocuous:

\begin{quoting}
\emph{I'm a woman, and I don't think you should vote for a woman just because she is a woman. Vote for her because you believe she deserves it.}
\end{quoting}

These comments were abridged from real comments posted by users in comments in the original article, as well as other online discussion forums discussing the issue (e.g., Reddit).

To ensure that the effects we observed were not path-dependent (i.e., if a discussion breaks down by chance because of a single user), we created eight separate ``universes'' for each condition \cite{salganik2006experimental}, for a total of 32 universes.
Each universe was seeded with the same comments, but were otherwise entirely independent.
Participants were randomized between universes within each condition.
Participants assigned to the same universe could see and respond to other participants who had commented prior, but not interact with participants from other universes.

\xhdr{Measuring discussion quality}
We evaluated discussion quality in two ways: if subsequent posts written exhibited trolling behavior, or if they contained more negative affect.
To evaluate whether a post was a troll post or not, two experts (including one of the authors) independently labeled posts as being troll or non-troll posts, blind to the experimental conditions, with disagreements resolved through discussion.
Both experts reviewed CNN.com's community guidelines \cite{cnnguidelines} for commenting -- posts that were offensive, irrelevant, or designed to elicit an angry response, whether intentional or not, were labeled as trolling.
To measure the negative affect of a post, we used LIWC \cite{pennebaker2001linguistic} (Vader \cite{hutto2014vader} gives similar results).

\begin{table}[tb]
\small
\centering
\ra{1.3}
\begin{tabular*}{\columnwidth}{@{\extracolsep{\fill} }l*{4}{D{.}{.}{1.2}}@{}}
\toprule
\emph{Fixed Effects} & \text{Coef.} & \text{SE} & \textit{z} \\
(Intercept) & -0.70^{***} & 0.17 & -4.23 \\
\textsc{NegMood} & 0.64^{**} & 0.24 & 2.66 \\
\textsc{NegContext} & 0.52^{*} & 0.23 & 2.38 \\
\textsc{NegMood} $\times$ \textsc{NegContext} & 0.41 & 0.33 & 1.23 \\
\midrule
\emph{Random Effects} & \text{Var.} & \text{SE} & \\
User & 0.41 & 0.64 & \\
\bottomrule
\end{tabular*}
\caption{A mixed effects logistic regression reveals a significant effect of both \textsc{NegMood} and \textsc{NegContext} on troll posts ($^{*}$: \textit{p}\textless0.05, $^{**}$: \textit{p}\textless0.01, $^{***}$: \textit{p}\textless0.001). In other words, both negative mood and the presence of initial troll posts increases the probability of trolling.}
\label{tab:exp_model}
\end{table}

\subsection{Results}

667 participants (40\% female, mean age  34.2, 54\% Democrat, 25\% Moderate, 21\% Republican) completed the experiment, with an average of 21 participants in each universe.
In aggregate, these workers contributed 791 posts (with an average of 37.8 words written per post) and 1392 votes.

\xhdr{Manipulation checks}
First we sought to verify that the \test did affect participants' mood. On average, participants in the \textsc{PosMood} condition obtained 11.2 out of 15 questions correct, performing above the stated ``average'' score of 8.
In contrast, participants in the \textsc{NegMood} condition answered only an average of 1.9 questions correctly, performing significantly worse (\textit{t}(594)=63.2, \textit{p}$<$0.001 using an unequal variances \textit{t}-test), and below the stated ``average''.
Correspondingly, the post-\test POMS questionnaire confirmed that participants in the \textsc{NegMood} condition experienced higher mood disturbance on all axes, with higher anger, confusion, depression, fatigue, and tension scores, and a lower vigor score (\textit{t}(534)$>$7.0, \textit{p}$<$0.001).
Total mood disturbance, where higher scores correspond to more negative mood, was 12.2 for participants in the \textsc{PosMood} condition (comparable to a baseline level of disturbance measured among athletes \cite{terry2000normative}), and 40.8 in the \textsc{NegMood} condition.
Thus, the \test put participants into a more negative mood.

Verifying that the initial posts in the \textsc{NegContext} condition were
perceived as being more troll-like than those in the \textsc{PosContext} condition, we found that the initial posts in the \textsc{NegContext} condition were less likely to be upvoted (36\% vs. 90\% upvoted for \textsc{PosContext}, \textit{t}(507)=15.7, \textit{p}$<$0.001).

\xhdr{Negative mood and negative context increase trolling behavior}
Table \ref{tab:exp_values} shows how the proportion of troll posts and negative affect
(measured as the proportion of negative words)
 differ in each condition.
The proportion of troll posts was highest in the (\textsc{NegMood}, \textsc{NegContext}) condition with 68\% troll posts, drops in both the (\textsc{NegMood}, \textsc{PosContext}) and (\textsc{PosMood}, \textsc{NegContext}) conditions with 47\% and 49\% each, and is lowest in the  (\textsc{PosMood}, \textsc{PosContext}) condition with 35\%.
For negative affect, we observe similar differences.

Fitting a mixed effects logistic regression model, with the two conditions as fixed effects, an interaction between the two conditions, user as a random effect, and whether a contributed post was trolling or not as the outcome variable, we do observe a significant effect of both \textsc{NegMood} and \textsc{NegContext} (\textit{p}$<$0.05) (Table \ref{tab:exp_model}).
These results confirm both H1 and H2, that negative mood and the discussion context (i.e., prior troll posts) increase a user's likelihood of trolling.
Negative mood increases the odds of trolling by 89\%, and the presence of prior troll posts increases the odds by 68\%.
A mixed model using MCMC revealed similar effects (\textit{p}$<$0.05), and controlling for universe, gender, age, or political affiliation also gave similar results.
Further, the effect of a post's position in the discussion on trolling was not significant, suggesting that trolling tends to persist in the discussion.

With the proportion of words with negative affect as the outcome variable, we observed a significant effect of \textsc{NegContext} (\textit{p}$<$0.05), but not of \textsc{NegMood}
-- such measures may not accurately capture types of trolling such as sarcasm or off-topic posting.
There was no significant effect of either factor on positive affect.

\xhdr{Examples of troll posts}
Contributed troll posts comprised a relatively wide range of antisocial behavior: from outright swearing (``What a dumb c***'') and personal attacks (``You're and idiot and one of the things that's wrong with this country.'') to veiled insults (``Hillary isn't half the man Bernie is!!! lol''), sarcasm (``You sound very white, and very male. Must be nice.''), and off-topic statements (``I think Ted Cruz has a very good chance of becoming president.'').
In contrast, non-troll posts tended to be more measured, regardless of whether they agreed with the article (``Honestly I agree too. I think too many people vote for someone who they identify with rather than someone who would be most qualified.'').

\xhdr{Other results}
We observed trends in the data.
Both conditions reduced the number of words written relative to the control condition: 44 words written in the (\textsc{PosMood}, \textsc{PosContext}) vs. 29 words written in the (\textsc{NegMood}, \textsc{NegContext}) condition.
Also, the percentage of upvotes on posts written by other users (i.e., excluding the initial seed posts) was lower: 79\% in the (\textsc{PosMood}, \textsc{PosContext}) condition vs. 75\% in the (\textsc{NegMood}, \textsc{NegContext}) condition.
While suggestive, neither effect was significant.

\xhdr{Discussion}
Why did \textsc{NegContext} and \textsc{NegMood} increase the rate of trolling?
Drawing on prior research explaining the mechanism of contagion \cite{wheeler1966toward}, participants may have an initial negative reaction to reading the article, but are unlikely to
bluntly externalize them because of self-control or environmental cues.
\textsc{NegContext} provides evidence
that others had similar reactions, making it more acceptable to also express them.
\textsc{NegMood} further accentuates any perceived negativity from reading the article and reduces self-inhibition \cite{leith1996bad}, making participants more likely to act out.

\xhdr{Limitations}
In this experiment, like prior work \cite{mcinnis2016one,sukumaran2011normative}, we recruited participants to participate in an online discussion, and required each to post at least one comment.
While this enables us isolate both mood and discussion context (which is difficult to control for in a live Reddit discussion for example) and further allows us to debrief participants afterwards, payment may alter the incentives to participate in the discussion.
Users also were commenting pseudonymously via randomly generated usernames, which may reduce overall comment quality \cite{kilner2005anonymity}.
Different initial posts may also elicit different subsequent posts.
While our analyses did not reveal significant effects of demographic factors, future work could further examine their impact on trolling.
For example, men may be more susceptible to trolling as they tend to be more aggressive \cite{berkowitz1993aggression}.
Anecdotally, several users who identified as Republican trolled the discussion with irrelevant mentions of Donald Trump (e.g., ``I'm a White man and I'm definitely voting for Donald Trump!!!'').
Understanding the effects of different types of trolling (e.g., swearing vs. sarcasm) and user motivations for such trolling (e.g., just to rile others up) also remains future work.
Last, different articles may be trolled to different extents \cite{gardiner2016trolls}, so we examine the effect of article topic in our subsequent analyses.

Overall, we find that both mood and discussion context significantly affect a user's likelihood of engaging in trolling behavior.
For such effects to be observable, a substantial proportion of the population must have been susceptible to trolling, rather than only a small fraction of
atypical
users -- suggesting that trolling can be generally induced.
But do these results generalize to real-world online discussions?
In the subsequent sections, we verify and extend our results with an analysis of CNN.com, a large online news discussion community.
After describing this dataset, we study how trolling behavior tracks known daily mood patterns, and how mood persists across multiple discussions.
We again find that the initial posts of discussions have a significant effect on subsequent posts, and study the impact of the volume and ordering of multiple troll posts on subsequent trolling.
Extending our analysis of discussion context to include the accompanying article's topic, we find that it too
mediates trolling behavior.

\section{Data: Introduction}
\label{sec:data}
CNN.com is a popular American news website where editors and journalists write articles on a variety of topics (e.g., politics and technology), which users can then discuss.
In addition to writing and replying to posts, users can {\em up-} and {\em down-vote}, as well as {\em flag} posts (typically for abuse or violations of the community guidelines \cite{cnnguidelines}).
Moderators can also {\em delete} posts or even {\em ban} users, in keeping with these guidelines.
Disqus, a commenting platform that hosted these discussions on CNN.com, provided us with a complete trace of user activity from December 2012 to August 2013, consisting of 865,248 users (20,197 banned), 16,470 discussions, and 16,500,603 posts, of which 571,662 (3.5\%) were flagged and 3,801,774 (23\%) were deleted.
Out of all flagged posts, 26\% were made by users with no prior record of flagging in previous discussions;
also, out of all users with flagged posts who authored at least ten posts, 40\% had less than
 3.5\% of their posts flagged (the baseline probability of
a random post being flagged on CNN).
These observations suggest that ordinary users are responsible for a significant amount of trolling behavior, and that many may have just been having a bad day.

In studying behavior on CNN.com, we consider two main units of analysis:
\begin{enumerate*}[label={\alph*)}]
\item a {\em discussion}, or all the posts that follow a given news article, and
\item a {\em sub-discussion}, or a top-level post and any replies to that post.
\end{enumerate*}
We make this distinction as discussions may reach thousands of posts, making it likely that users may post in a discussion without reading any previous responses.
In contrast, a sub-discussion necessarily involves replying to a previous post, and would allow us to better study the effects of people reading and responding to each other.

In our subsequent analyses, we filter banned users (of which many tend to be clearly identifiable trolls \cite{cheng2015antisocial}), as well as any users who had all of their posts deleted, as we are primarily interested in studying the effects of mood and discussion context on the general population.

We use flagged posts (posts that CNN.com users marked for violating community guidelines) as our primary measure of trolling behavior.
In contrast, moderator deletions are typically incomplete: moderators miss some legitimate troll behavior and tend to delete entire discussions as opposed to individual posts.
Likewise, written negative affect misses sarcasm and other trolling behaviors that do not involve common negative words, and downvoting may simply indicate disagreement.
To validate this approach, two experts
(including one of the authors)
labeled 500 posts (250 flagged) sampled at random, blind to whether each post was flagged, using the same criteria for trolling as for the experiment.
Comparing the expert labels with post flags from the dataset, we obtained a precision of 0.66 and recall of 0.94, suggesting that while some troll posts remain unflagged, almost all flagged posts are troll posts.
In other words, while instances of trolling behavior go unnoticed (or are ignored), when a post is flagged, it is highly likely that trolling behavior did occur.
So, we use flagged posts as a primary estimate of trolling behavior in our analyses, complementing our analysis with other signals such as negative affect and downvotes.
These signals are correlated: flagged posts are more likely than non-flagged posts to have greater negative affect (3.7\% vs. 3.4\% of words, Cohen's \textit{d}=0.06, \textit{t}=40, \textit{p}$<$0.001), be downvoted (58\% vs. 30\% of votes, \textit{d}=0.76, \textit{t}=531, \textit{p}$<$0.001), or be deleted by a moderator (79\% vs. 21\% of posts, \textit{d}=1.4, \textit{t}=1050, \textit{p}$<$0.001).

\section{Data: Understanding Mood}
\label{sec:mood}
\begin{figure*}[tb]
        \centering
        \begin{subfigure}[b]{0.33\textwidth}
                \includegraphics[width=\textwidth]{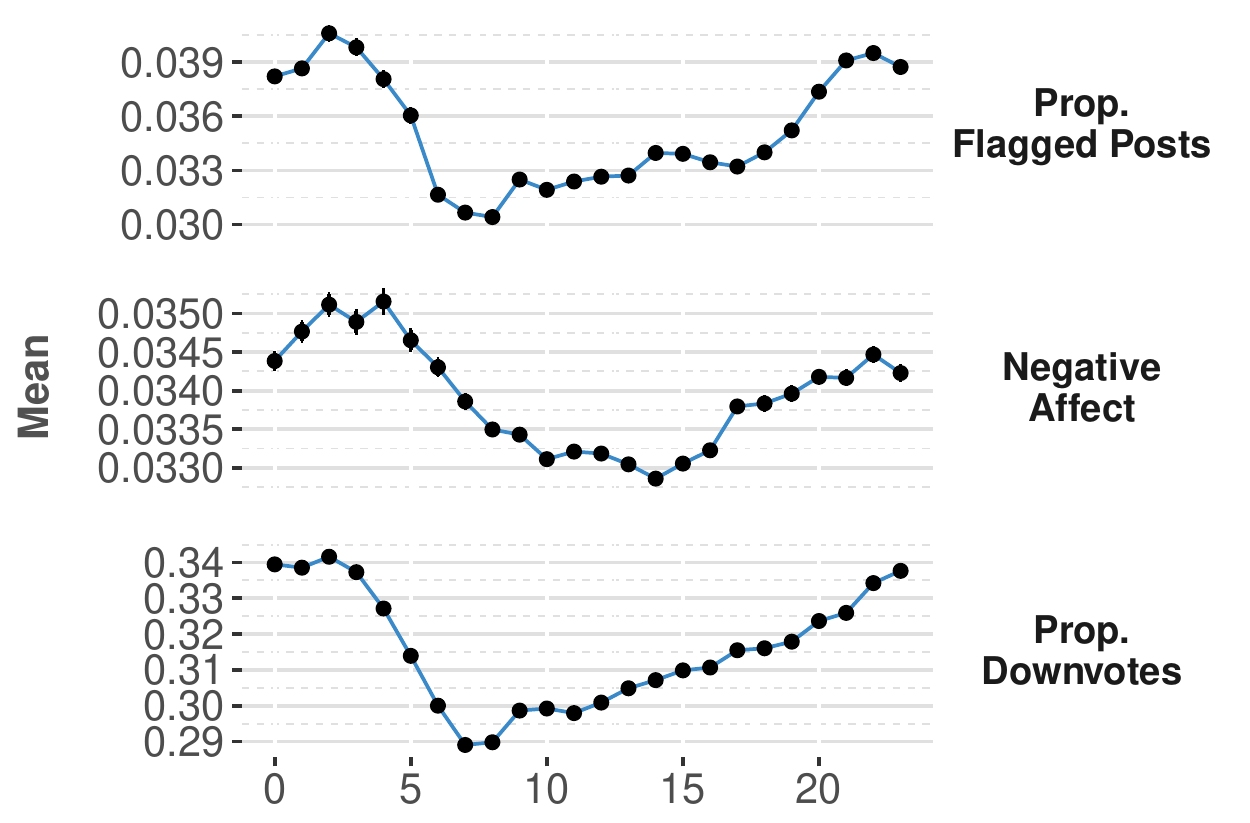}
                \caption{Time of Day (EDT)}
                \label{fig:seasonality_1}
        \end{subfigure}
        \begin{subfigure}[b]{0.33\textwidth}
                \includegraphics[width=\textwidth]{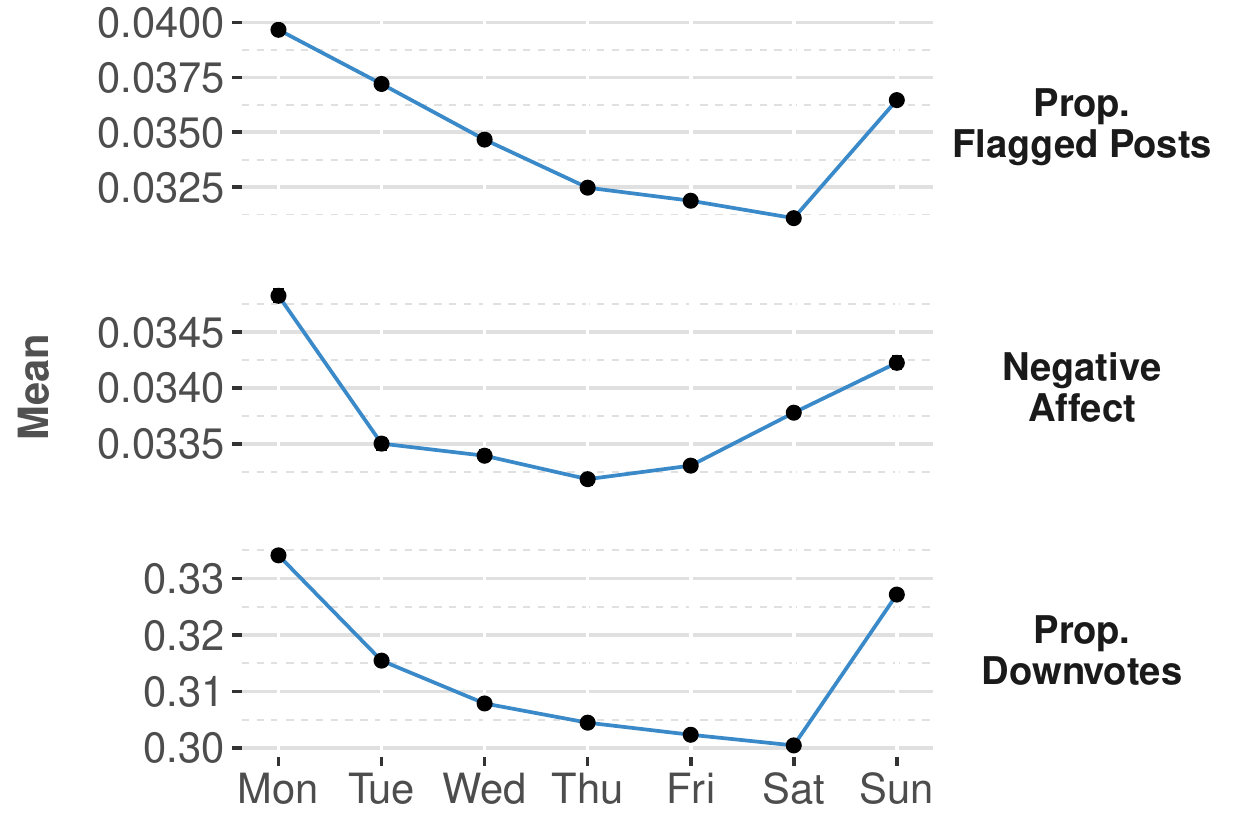}
                \caption{Day of Week}
                \label{fig:seasonality_2}
        \end{subfigure}
        \begin{subfigure}[b]{0.33\textwidth}
                \includegraphics[width=\textwidth]{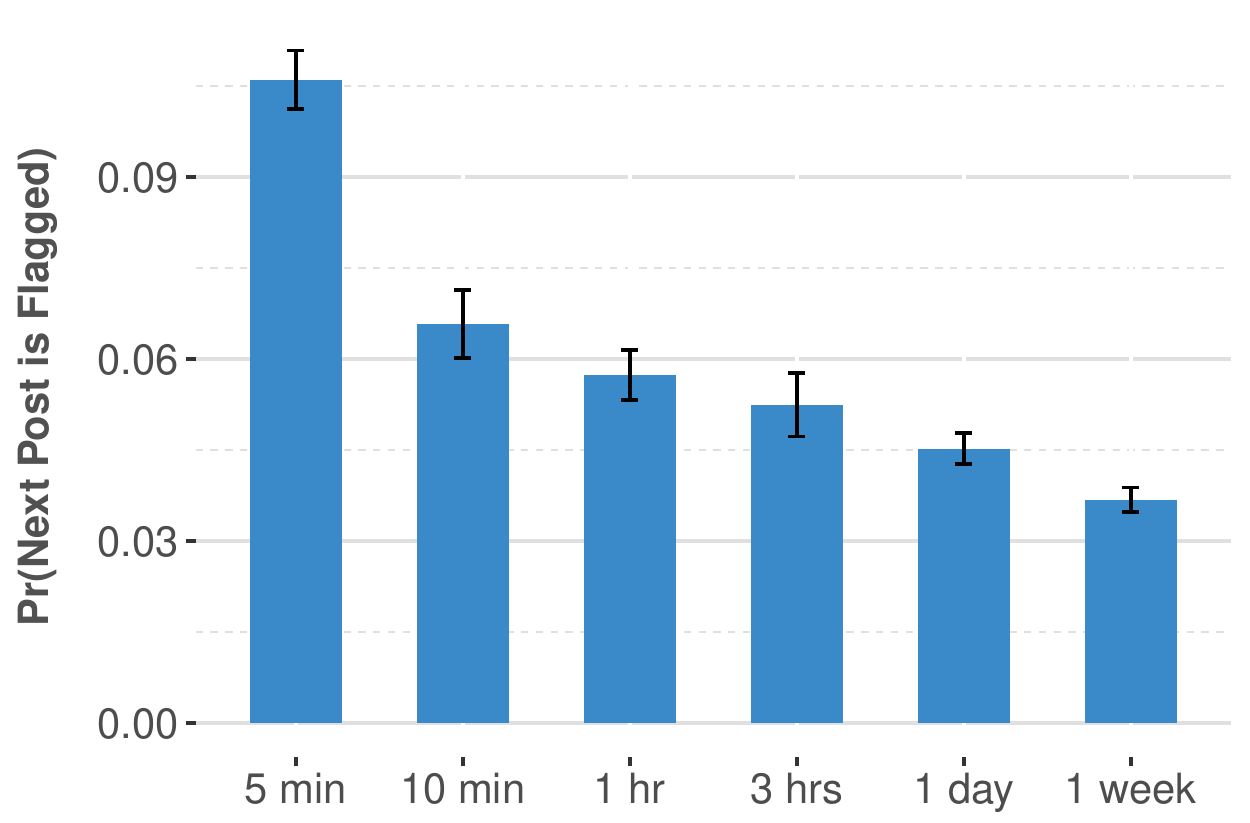}
                \caption{Time Since Last Post}
                \label{fig:seasonality_3}
        \end{subfigure}
        \caption{Like negative mood, indicators of trolling peak (a) late at night, and (b) early in the work week, supporting a relation between mood and trolling. Further, (c) the shorter the time between a user's subsequent posts in unrelated discussions,
        where the first post is flagged, the more likely the second will also be flagged, suggesting that negative mood may persist for some time.
        }
        \label{fig:seasonality}
\end{figure*}

\begin{figure}[t]
  \includegraphics[width=\columnwidth]{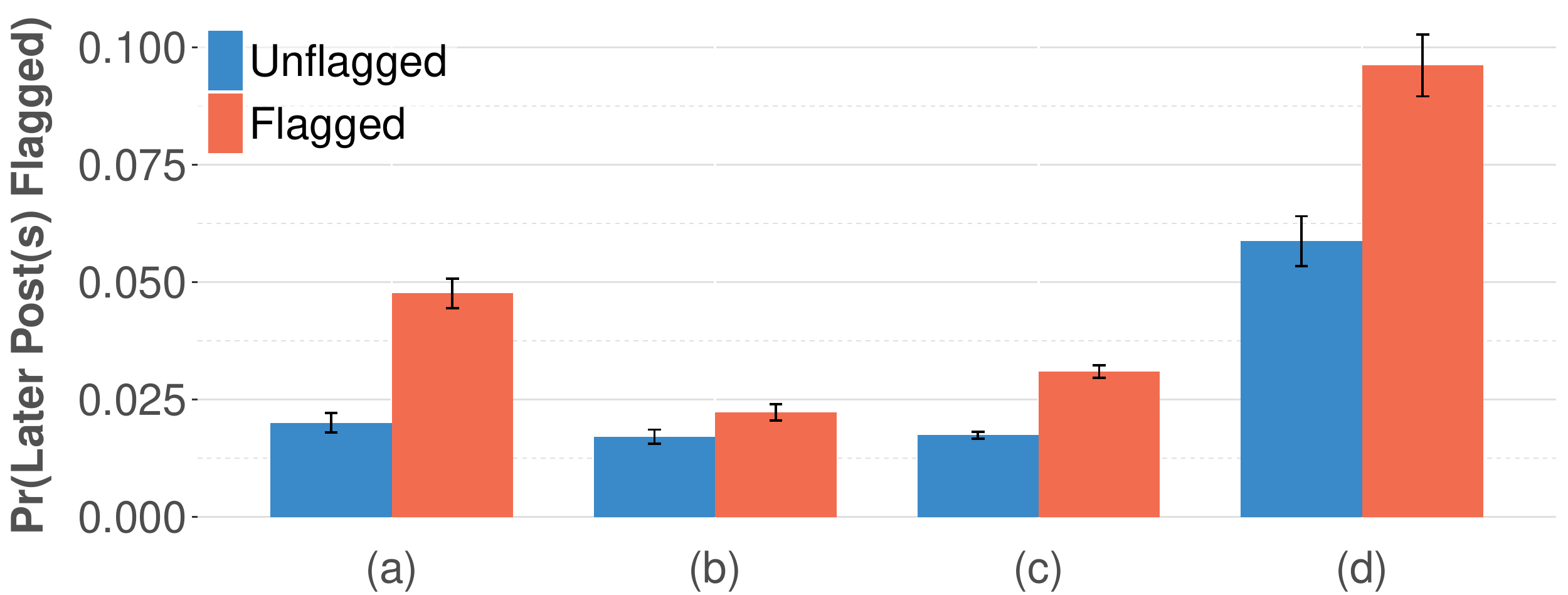}
  \caption{Suggesting that negative mood may persist across discussions, users with no prior history of flagged posts, who either (a) make a post in a prior unrelated discussion that is flagged, or (b) simply participates in a sub-discussion in a prior discussion with at least one flagged post, without themselves being flagged, are more likely to be subsequently flagged in the next discussion they participate in. Demonstrating the effect of discussion context, (c) discussions that begin with a flagged post are more likely to have a greater proportion of flagged posts by other users later on, as do (d)~sub-discussions that begin with a flagged post.}
  \label{fig:replication}
\end{figure}

In the earlier experiment, we showed that bad mood increases the probability of trolling.
In this section, using large-scale and longitudinal observational data, we verify and expand on this result.
While we cannot measure mood directly, we can study its known correlates.
Seasonality influences mood \cite{golder2011diurnal}, so we study how trolling behavior also changes with the time of day or day of week.
Aggression can linger beyond an initial unpleasant event \cite{jones1978air}, thus we also study how trolling behavior persists as a user participates in multiple discussions.

\subsection{Happy in the day, sad at night}
Prior work that studied changes in linguistic affect on Twitter demonstrated that mood changes with the time of the day, and with the
day of the week -- positive affect peaks in the morning, and during weekends \cite{golder2011diurnal}.
If mood changes with time, could trolling be similarly affected?
Are people more likely to troll later in the day, and on weekdays?
To evaluate the impact of the time of day or day of week on mood and trolling behavior, we track several measures that may indicate troll-like behavior:
\begin{enumerate*}[label={\alph*)}]
\item the proportion of flagged posts (or posts reported by other users as being abusive),
\item negative affect, and
\item the proportion of downvotes on posts (or the average fraction of downvotes on posts that received at least one vote).
\end{enumerate*}

Figures \ref{fig:seasonality_1} and \ref{fig:seasonality_2} show how each of these measures changes with the time of day and day of week, respectively, across all posts.
Our findings corroborate prior work -- the proportion of flagged posts, negative affect, and the proportion of downvotes are all lowest in the morning, and highest in the evening, aligning with when mood is worst \cite{golder2011diurnal}.
These measures also peak on Monday (the start of the work week in the US).

Still, trolls may simply wake up later than normal users, or post on different days.
To understand how the time of day and day of week affect the same user, we compare these measures for the same user in two different time periods: from 6 am to 12 pm and from 11 pm to 5 am, and on two different days: Monday and Friday (i.e., early or late in the work week).
A paired \textit{t}-test reveals a small, but significant increase in negative behavior between 11 pm and 5 am (flagged posts: 4.1\% vs. 4.3\%, \textit{d}=0.01, \textit{t}(106300)=2.79, \textit{p}$<$0.01; negative affect: 3.3\% vs. 3.4\%, \textit{t}(106220)=3.44, \textit{d}=0.01, \textit{p}$<$0.01; downvotes: 20.6\% vs. 21.4\%, \textit{d}=0.02, \textit{t}(26390)=2.46, \textit{p}$<$0.05).
Posts made on Monday also show more negative behavior than posts made on Friday (\textit{d}$\ge$0.02, \textit{t}$>$2.5, \textit{p}$<$0.05).
While these effects may also be influenced by the type of news that gets posted at specific times or days, limiting our analysis to just news articles categorized as ``US'' or ``World'', the two largest sections, we continue to observe similar results.

Thus, even without direct user mood measurements, patterns of trolling behavior
correspond predictably with mood.

\subsection{Anger begets more anger}
Negative mood can persist beyond the events that brought about those feelings \cite{keltner1993beyond}.
If trolling is dependent on mood, we may be able to observe the aftermath of user outbursts, where negative mood might spill over from prior discussions into subsequent, unrelated ones, just as our experiment showed that negative mood that resulted from doing poorly on a \test affected later commenting in a discussion.
Further, we may also differentiate the effects that stem from actively engaging in negative behavior in the past, versus simply being exposed to negative behavior.
Correspondingly, we ask two questions, and answer them in turn.
First,
\begin{enumerate*}[label={\alph*)}]
\item if a user wrote a troll post in a prior discussion, how does that affect their probability of trolling in a subsequent, unrelated discussion? At the same time, we might also observe indirect effects of trolling:
\item if a user participated in a discussion where trolling occurred, but did not engage in trolling behavior themselves, how does that affect their probability of trolling in a subsequent, unrelated discussion?
\end{enumerate*}

To answer the former, for a given discussion, we sample two users at random, where one had a post which was flagged, and where one had a post which was not flagged.
We ensure that these two users made at least one post prior to participating in the discussion, and match users on the total number of posts they wrote prior to the discussion.
As we are interested in these effects on ordinary users, we also ensure that neither of these users have had any of their posts flagged in the past.
We then compare the likelihood of each user's next post in a new discussion also being flagged.
We find that users who had a post flagged in a prior discussion were twice as likely to troll in their next post in a different discussion (4.6\% vs. 2.1\%, \textit{d}=0.14, \textit{t}(4641)=6.8, \textit{p}$<$0.001) (Figure \ref{fig:replication}a).
We obtain similar results even when requiring these users to also have no prior deleted posts or longer histories (e.g., if they have written at least five posts prior to the discussion).

Next, we examine the indirect effect of participating in a ``bad'' discussion, even when the user does not directly engage in trolling behavior.
We again sample two users from the same discussion, but where each user participated in a different sub-discussion: one sub-discussion had at least one other post by another user flagged, and the other sub-discussion had no flagged posts.
Again, we match users on the number of posts they wrote in the past, and ensure that these users have no prior flagged posts (including in the sampled discussions).
We then compare the likelihood of each user's next post in a new discussion being flagged.
Here, we also find that users who participated in a prior discussion with at least one flagged post were significantly more likely to subsequently author a post in an new discussion that would be flagged (Figure \ref{fig:replication}b).
However, this effect is significantly weaker (2.2\% vs. 1.7\%, \textit{d}=0.04, \textit{t}(7321)=2.7, \textit{p}$<$0.01).

Thus, both trolling in a past discussion, as well as participating in a discussion where trolling occurred, can affect whether a user trolls in the future discussion. These results suggest that negative mood can persist and transmit trolling norms and behavior across multiple discussions, where there is no similar context to draw on.
As none of the users we analyzed had prior flagged posts, this effect is unlikely to arise simply because some users were just trolls in general.

\subsection{Time heals all wounds}
One typical anger management strategy is to use a ``time-out'' to calm down \cite{kendall1975timeout}.
Thus, could we minimize negative mood carrying over to new discussions by having users wait longer before making new posts?
Assuming that a user is in a negative mood (as indicated by writing a post that is flagged), the time elapsed until the user's next post may correlate with the likelihood of subsequent trolling.
In other words, we might expect that the longer time the time between posts, the greater the temporal distance from the origin of the negative mood, and hence the lower the likelihood of trolling.

Figure \ref{fig:seasonality_3} shows how the probability of a user's next post being flagged changes with the time since that user's last post, assuming that the previous post was flagged.
So as not to confuse the effects of the initial post's discussion context, we ensure that the user's next post is made {\em in a new discussion with different other users}.
The probability of being flagged is high when the time between these two subsequent posts is short (five minutes or less), suggesting that a user might still be in a negative mood persisting from the initial post.
As more time passes, even just ten minutes, the probability of being flagged gradually decreases.
Nonetheless, users with better impulse control may wait longer before posting again if they are angry, and isolating this effect would be future work.
Our findings here lend credence to the rate-limiting of posts that some forums have introduced \cite{discourse2013topic}.

\section{Data: Understanding Discussion Context}
\label{sec:context}
\begin{figure*}[tb]
        \centering
        \begin{subfigure}[b]{0.33\textwidth}
                \includegraphics[width=\textwidth]{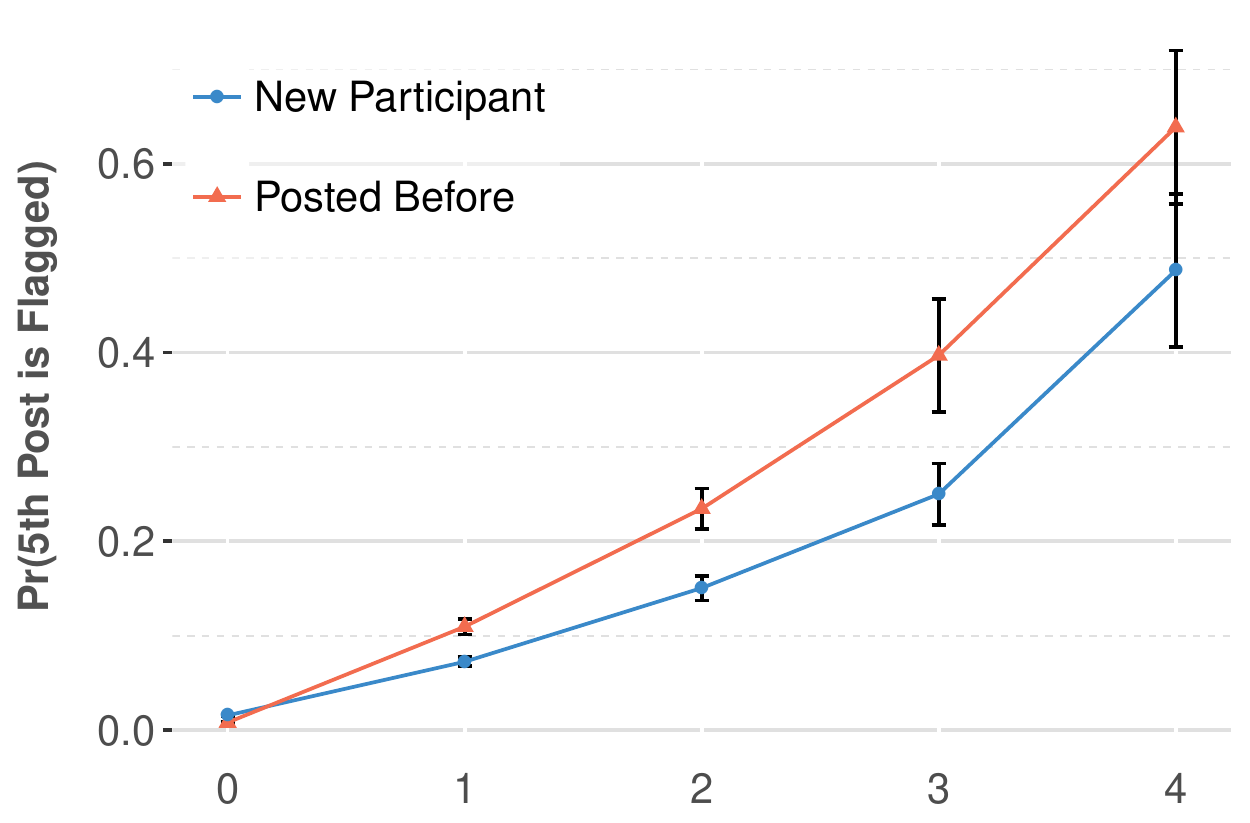}
                \caption{\# of Prior Flagged Posts}
                \label{fig:context_1}
        \end{subfigure}
        \begin{subfigure}[b]{0.33\textwidth}
                \includegraphics[width=\textwidth]{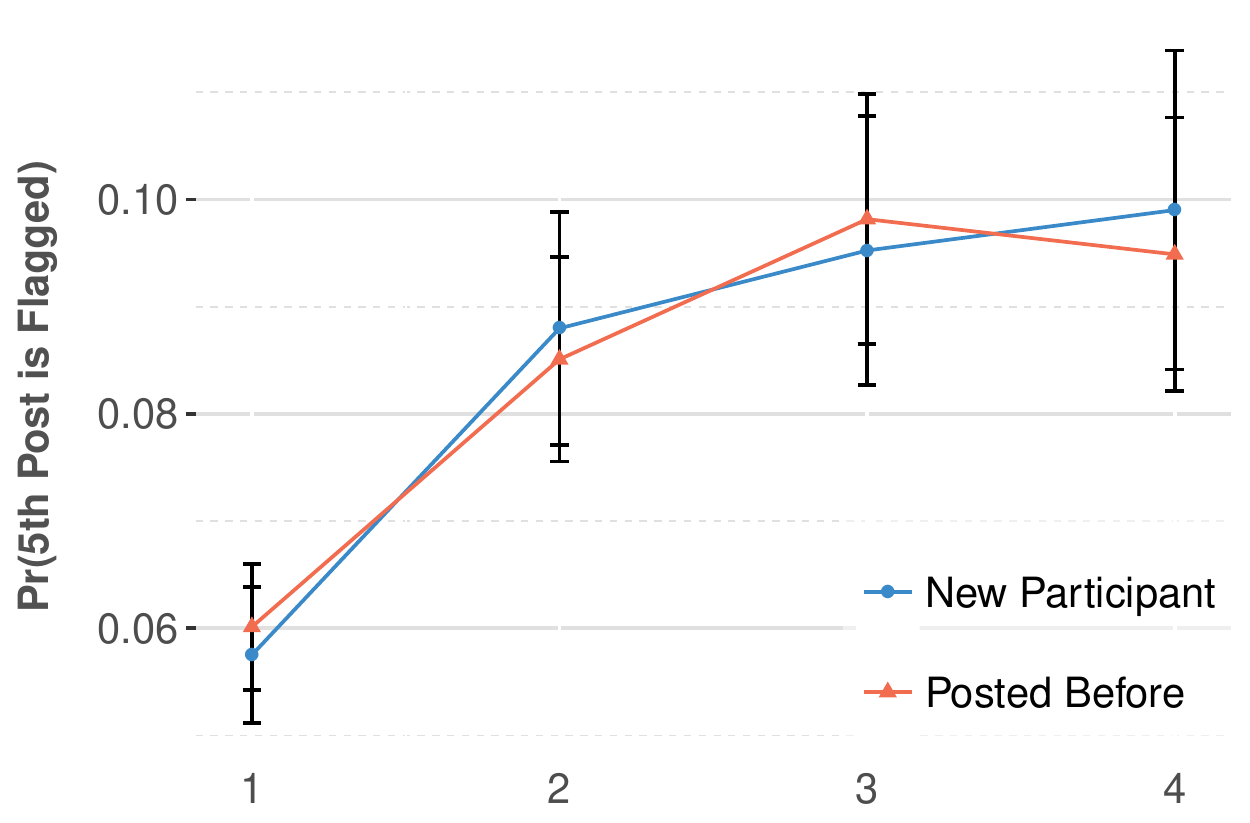}
                \caption{Position of Flagged Post}
                \label{fig:context_2}
        \end{subfigure}
        \begin{subfigure}[b]{0.33\textwidth}
                \includegraphics[width=\textwidth]{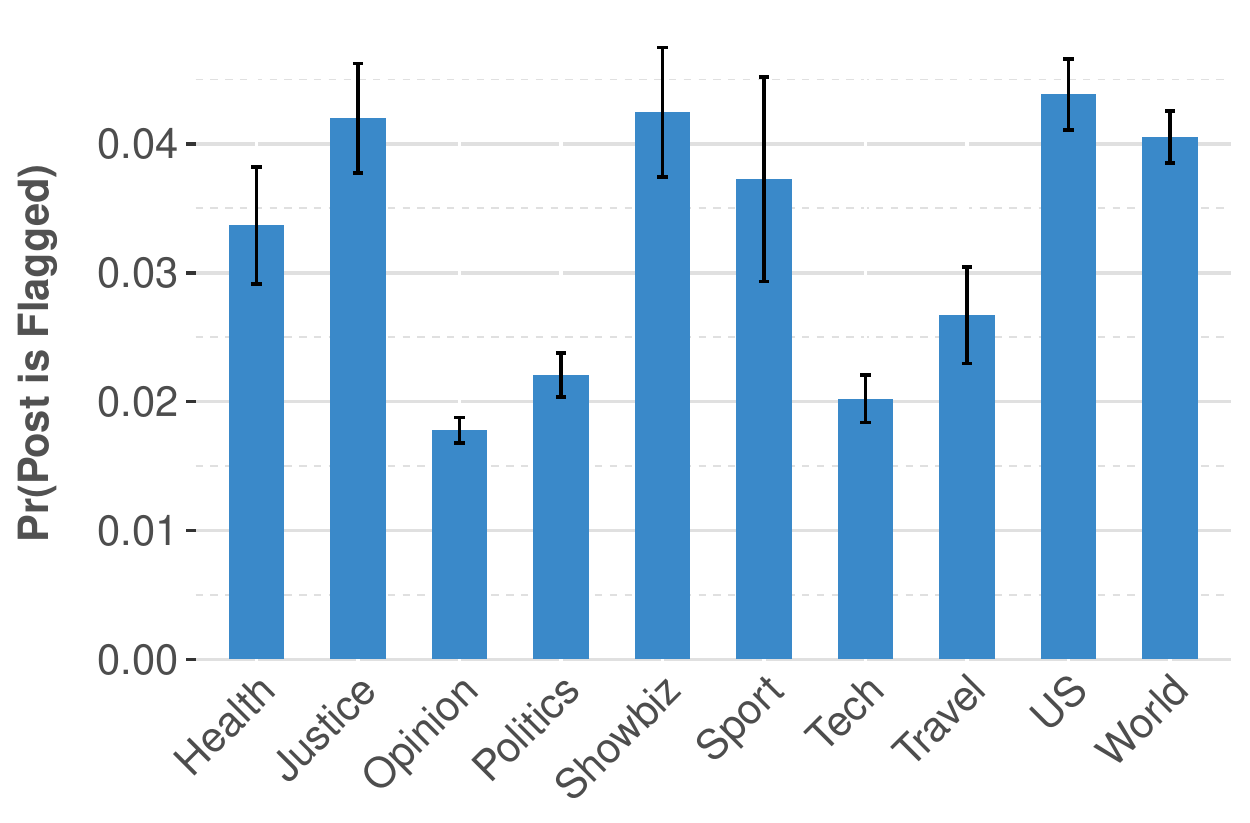}
                \caption{Flagged Posts by Topic}
                \label{fig:context_3}
        \end{subfigure}
        \caption{In discussions with at least five posts, (a) the probability that a post is flagged monotonically increases with the number of prior flagged posts in the discussion. (b) If only one of the first four posts was flagged, the fifth post is more likely to be flagged if that flagged post is closer in position. (c) The topic of a discussion also influences the probability of a post being flagged.}
        \label{fig:context}
\end{figure*}

From our experiment, we identified mood and discussion context as influencing trolling.
The previous section verified and extended our results on mood; in this section, we do the same for discussion context.
In particular, we show that posts are more likely to be flagged if others' prior posts were also flagged.
Further, the number and ordering of flagged posts in a discussion affects the probability of subsequent trolling, as does the topic of the discussion.

\subsection{``FirST!!1''}
How strongly do the initial posts to a discussion affect the likelihood of subsequent posts to troll?
To measure the effect of the initial posts on subsequent discussions, we first identified discussions of at least 20 posts, separating them into those with their first post flagged and those without their first post flagged.
We then used propensity score matching to create matched pairs of discussions where the topic of the article, the day of week the article was posted, and the total number of posts are controlled for \cite{rosenbaum1983central}.
Thus, we end up with pairs of discussions on the same topic, started on the same day of the week, and with similar popularity, but where one discussion had its first post flagged, while the other did not.
We then compare the probability of the subsequent posts in the discussion being flagged.
As we were interested in the impact of the initial post on other ordinary users, we excluded any posts written by the user who made the initial post, posts by users who replied (directly or indirectly) to that post, and posts by users with prior flagged or deleted posts in previous discussions.

After an initial flagged post, we find that subsequent posts by other users were more likely to be flagged, than if the initial post was not flagged (3.1\% vs. 1.7\%, \textit{d}=0.32, \textit{t}(1545)=9.1, \textit{p}$<$0.001) (Figure \ref{fig:replication}c).
This difference remains significant even when only considering posts made in the second half of a discussion (2.1\% vs. 1.3\%, \textit{d}=0.19, \textit{t}(1545)=5.4, \textit{p}$<$0.001).
Comparing discussions where the first three posts were all flagged to those where none of these posts were flagged (similar to \textsc{NegContext} vs. \textsc{PosContext} in our experiment), the gap widens (7.1\% vs. 1.7\%, \textit{d}=0.61, \textit{t}(113)=4.6, \textit{p}$<$0.001).

Nonetheless, as these different discussions were on different articles, some articles, even within the same topic, may have been more inflammatory, increasing the overall rate of flagging.
To control for the article being discussed, we also look at sub-discussions (a top-level post and all of its replies) within the same discussion.
Sub-discussions tend to be closer to actual conversations between users as each subsequent post is an explicit reply to another post in the chain, as opposed to considering the discussion as a whole where users can simply leave a comment without reading or responding to anyone else.
From each discussion we select two sub-discussions at random, where one sub-discussion's top-level post was flagged, and where the other's was not, and only considered posts not written by the users who started these sub-discussions.
Again, we find that sub-discussions whose top-level posts were flagged were significantly more likely to result in more flagging later in that sub-discussion (9.6\% vs. 5.9\%, \textit{d}=0.16, \textit{t}(501)=3.9, \textit{p}$<$0.001) (Figure \ref{fig:replication}d).

Altogether, these results suggest that the initial posts in a discussion set a strong, lasting precedent for later trolling.

\subsection{From bad to worse: sequences of trolling}
By analyzing the volume and ordering of troll posts in a discussion, we can better understand how discussion context and trolling behavior interact.
Here, we study sub-discussions at least five posts in length, and separately consider posts written by users new to the sub-discussion and posts written by users who have posted before in the sub-discussion to control for the impact of having already participated in the discussion.

Do more troll posts increase the likelihood of future troll posts?
Figure \ref{fig:context_1} shows that
as the number of flagged posts among the first four posts increases,
the probability that the fifth post is also flagged increases monotonically.
With no prior flagged posts, the chance of the fifth post by a new user to the sub-discussion being flagged is just 2\%; with one other flagged post, this jumps to 7\%; with four flagged posts, the odds of the fifth post also being flagged are almost one to one (49\%).
These pairwise differences are all significant with a Holm correction ($\chi^2$(1)$>$7.6, \textit{p}$<$0.01).
We observe similar trends for users new to the sub-discussion, as well as users that had posted previously, with the latter group of users more likely to be subsequently flagged.

Further, does a troll post made later in a discussion, and closer to where a user's post will show up, have a greater impact than a troll post made earlier on?
Here, we look at discussions of at least five posts where there was exactly one flagged post among the first four, and where that flagged post was not written by the fifth post's author.
In Figure~\ref{fig:context_2}, the closer in position the flagged post is to the fifth post, the more likely that post is to be flagged. For both groups of users, the fifth post in a discussion is more likely to be flagged if the fourth post was flagged, as opposed to the first ($\chi^2$(1)$>$6.9, \textit{p}$<$0.01).

Beyond the presence of troll posts, their conspicuousness in discussions substantially affects if new discussants troll as well.
These findings,
together
with our previous results showing how simply participating in a previous discussion having a flagged post raises the likelihood of future trolling behavior, support H3: that trolling behavior spreads from user to user.

\subsection{Hot-button issues push users' buttons?}
How does the subject of a discussion affect the rate of trolling?
Controversial topics (e.g., gender, GMOs, race, religion, or war) may divide a community \cite{levine1943learning}, and thus lead to more trolling.
Figure \ref{fig:context_3} shows the average rate of flagged posts of articles belonging to different sections of CNN.com.

Post flagging is more frequent in the health, justice, showbiz, sport, US, and world sections (near 4\%), and less frequent in the opinion, politics, tech, and travel sections (near 2\%).
Flagging may be more common in the health, justice, US, and world sections because these sections tend to cover controversial issues: a linear regression unigram model using the titles of articles to predict the proportion of flagged posts revealed that ``Nidal'' and ``Hasan'' (the perpetrator of the 2009 Fort Hood shooting) were among the most predictive words in the justice section.
For the showbiz and sport sections, inter-group conflict may have a strong effect (e.g., fans of opposing teams) \cite{sherif1961intergroup}.
Though political issues in the US may appear polarizing, the politics section has one of the lowest rates of post flagging, similar to tech.
Still, a deeper analysis of the interplay of these factors (e.g., personal values, group membership, and topic) with trolling remains future work.

The relatively large variation here suggests that the topic of a discussion influences the baseline rate of trolling, where hot-button topics spark more troll posts.

\subsection{Summary}
Through experimentation and data analysis, we find that situational factors such as mood and discussion context can induce trolling behavior, answering our main research question (RQ). Bad mood induces trolling, and trolling, like mood, varies with time of day and day of week; bad mood may also persist across discussions, but its effect  diminishes with time.
Prior troll posts in a discussion increase the likelihood of future troll posts (with an additive effect the more troll posts there are), as do more controversial topics of discussion.

\section{A Model of How Trolling Spreads}
\label{sec:model}
\begin{table}[tb]
\small
\centering
\ra{1.3}
\begin{tabular*}{\columnwidth}{@{\extracolsep{\fill}}ll}\toprule
  \textbf{Feature Set} & AUC \\
  \hline
  \multicolumn{2}{c}{\textit{Mood}} \\
  Seasonality (31) & 0.53 \\

  Recent User History (4) & 0.60 \\
  \hline
  \multicolumn{2}{c}{\textit{Discussion Context}} \\
  Previous Posts (15) & 0.74 \\

  Article Topic (13) & 0.58 \\

  \hline
    \multicolumn{2}{c}{\textit{User-specific}} \\

  Overall User History (2) & 0.66 \\

  User ID (45895) & 0.66 \\
  \hline
    \multicolumn{2}{c}{\textit{Combined}} \\
    Previous Posts + Recent User History (19) & 0.77 \\
  All Features & 0.78 \\
\bottomrule
\end{tabular*}
\caption{In predicting trolling in a discussion, features relating to the discussion's context are most informative, followed by user-specific and mood features. This suggests that while some users are inherently more likely to troll, the context of a discussion plays a greater role in whether trolling actually occurs. The number of binary features is in parentheses.}
\label{tab:performance}
\end{table}

Thus far, our investigation sought to understand whether ordinary users engage in trolling behavior.
In contrast, prior work suggested that trolling is largely driven by a small population of trolls (i.e., by intrinsic characteristics such as personality), and our evidence suggests complementary hypotheses -- that mood and discussion context also affect trolling behavior.
In this section, we construct a combined predictive model to understand the relative strengths of each explanation.

We model each explanation through features in the CNN.com dataset.
First, the impact of mood on trolling behavior can be modeled indirectly using \emph{seasonality}, as expressed through time of day and day of week; and a user's \emph{recent posting history} (outside of the current discussion), in terms of the time elapsed since the last post and whether the user's previous post was flagged.
Second, the effect of discussion context can be modeled using the \emph{previous posts} that precede a user's in a discussion (whether any of the previous five posts in the discussion were flagged, and if they were written by the same user); and the \emph{topic} of discussion (e.g., politics).
Third, to evaluate if trolling may be innate, we use a user's \emph{User ID} to learn each user's base propensity to troll, and the user's \emph{overall history} of prior trolling (the total number and proportion of flagged posts accumulated).

Our prediction task is to guess whether a user will write a post that will get flagged, given features relating to the discussion or user.
We sampled posts from discussions at random (\textit{N}=116,026), and balance the set of users whose posts are later flagged and users whose posts are not flagged, so that random guessing results in 50\% accuracy.
To understand trolling behavior across all users, this analysis was not restricted to users who did not have their posts previously flagged.
We use a logistic regression classifier, one-hot encoding features (e.g., time of day) as appropriate.
A random forest classifier gives empirically similar results.

Our results suggest that trolling is better explained as situational (i.e., a result of the user's environment) than as innate (i.e., an inherent trait).
Table \ref{tab:performance} describes performance on this prediction task for different sets of features.
Features relating to discussion context perform best (AUC=0.74), hinting that context alone is sufficient in predicting trolling behavior; the individually most predictive feature was whether the previous post in the discussion was flagged.
Discussion topic was somewhat informative (0.58), with the most predictive feature being if the post was in the opinion section.
In the experiment, mood produced a stronger effect than discussion context.
However, here we cannot measure mood directly, so its feature sets (seasonality and recent user history) were weaker (0.60 and 0.53 respectively).
Most predictive was if the user's last post in a different discussion was flagged, and if the post was written on Friday.
Modeling each user's probability of trolling individually, or by measuring all flagged posts over their lifetime was moderately predictive (0.66 in either case).
Further, user features do not improve performance beyond the using just the discussion context and a user's recent history.
Combining previous posts with recent history (0.77) resulted in performance nearly as good as including all features (0.78).
We continue to observe strong performance when restricting our analysis only to posts by users new to a discussion (0.75), or to users with no prior record of reported or deleted posts (0.70).
In the latter case, it is difficult to detect trolling behavior without discussion context features ($<$0.56).

Overall, we find that the context in which a post is made is a strong predictor of a user later trolling, beyond their intrinsic propensity to troll.
A user's
recent
posting history is also predictive, suggesting that mood carries over from previous discussions, and that past trolling predicts future trolling.

\section{Discussion}
\label{sec:discussion}
\begin{figure}[tb]
\includegraphics[width=\columnwidth]{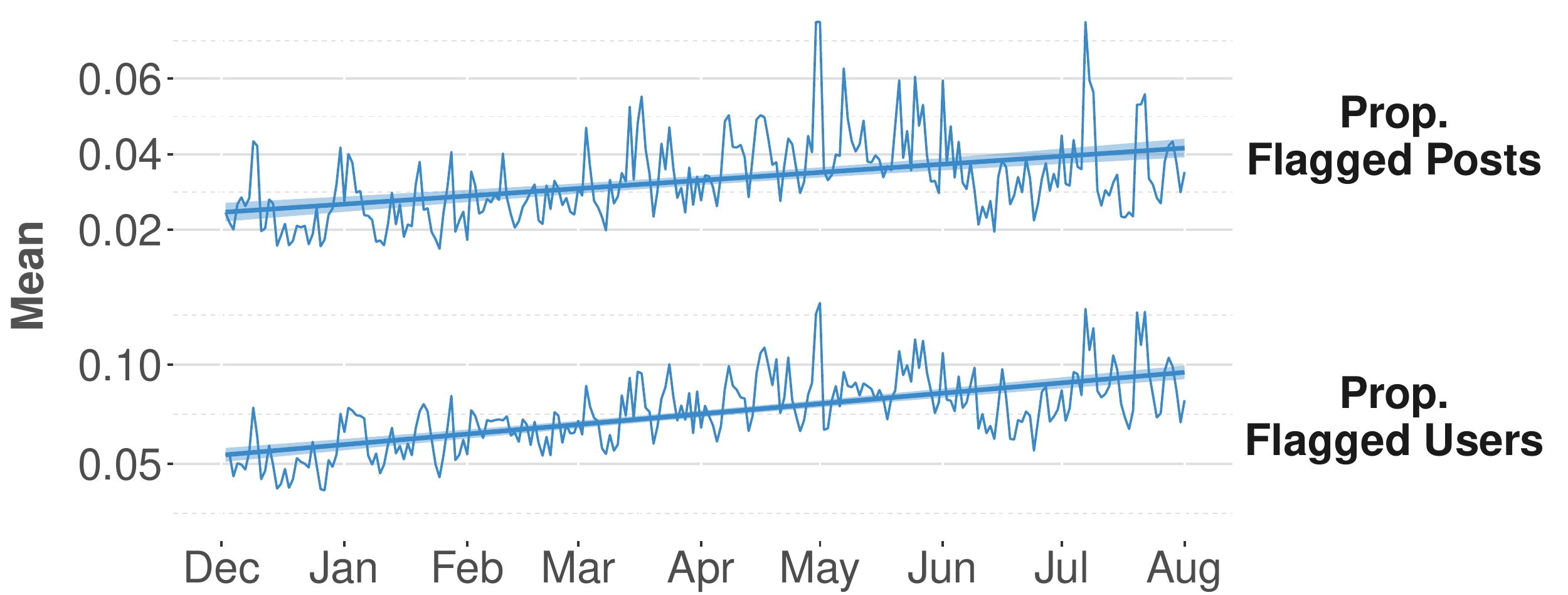}
\caption{On CNN.com, the proportion of flagged posts, as well as users with flagged posts, is increasing over time, suggesting that trolling behavior can spread and be reinforced.}
\label{fig:overtime}
\end{figure}

While prior work suggests that some users may be born trolls and innately more likely to troll others, our results show that ordinary users will also troll when mood and discussion context prompt such behavior.

\subsection{The spread of negativity}
If trolling behavior can be induced, and can carry over from previous discussions, could such behavior cascade and lead to the community worsening overall over time?
Figure \ref{fig:overtime} shows that on CNN.com, the proportion of flagged posts and proportion of users with flagged posts are rising over time.
These upward trends suggest that trolling behavior is becoming more common, and that a growing fraction of users are engaging in such behavior.
Comparing posts made in the first half and second half of the CNN.com dataset, the proportion of flagged posts and proportion of users with flagged posts increased (0.03 vs. 0.04 and 0.09 vs. 0.12, \textit{p}$<$0.001).
There may be several explanations for this (e.g., that users joining later are more susceptible to trolling), but our findings, together with prior work showing that negative norms can be reinforced \cite{willer2009false} and that downvoted users go on to downvote others \cite{cheng2014community}, suggest that negative behavior can persist in and permeate a community when left unchecked.

\subsection{Designing better discussion platforms}
The continuing endurance of the idea that trolling is innate may be explained using the fundamental attribution error \cite{ross1977intuitive}: people tend to attribute a person's behavior to their internal characteristics rather than external factors -- for example, interpreting snarky remarks as resulting from general mean-spiritedness (i.e., their disposition), rather than a bad day (i.e., the situation that may have led to such behavior).
This line of reasoning may lead communities to incorrectly conclude that trolling is caused by people who are unquestionably trolls, and that trolling can be eradicated by banning these users.
However, not only are some banned users likely to be ordinary users just having a bad day, but such an approach also does little to curb such situational trolling, which many ordinary users may be susceptible to.
How might we design discussion platforms that minimize the spread of trolling behavior?

Inferring mood through recent posting behavior (e.g., if a user just participated in a heated debate) or other behavioral traces such as keystroke movements \cite{kolakowska2013review}, and selectively enforcing measures such as post rate-limiting \cite{discourse2013topic} may discourage users from posting in the heat of the moment.
Allowing users to retract recently posted comments may help minimize regret \cite{wang2011regretted}.
Alternatively, reducing other sources of user frustration (e.g., poor interface design or slow loading times \cite{ceaparu2004determining}) may further temper aggression.

Altering the context of a discussion (e.g., by hiding troll comments and prioritizing constructive ones) may increase the perception of civility, making users less likely to follow suit in trolling.
To this end, one solution is to rank comments using user feedback, typically by allowing users to up- and downvote content, which reduces the likelihood of subsequent users encountering downvoted content.
But though this approach is scalable, downvoting can cause users to post worse comments, perpetuating a negative feedback loop \cite{cheng2014community}.
Selectively exposing feedback, where positive signals are public and negative signals are hidden, may enable context to be altered without adversely affecting user behavior.
Community norms can also influence a discussion's context: reminders of ethical standards or past moral actions (e.g., if users had to sign a ``no trolling'' pledge before joining a community) can also increase future moral behavior \cite{mazar2008dishonesty,nathan2016posting}.

\subsection{Limitations and future work}
Though our results do suggest the overall effect of mood on trolling behavior, a more nuanced understanding of this relation should require improved signals of mood (e.g., by using behavioral traces as described earlier).
Models of discussions that account for the reply structure \cite{Backstrom:ProceedingsOfWsdm:2013}, changes in sentiment \cite{Wang:ProceedingsOfTheAcl:}, and the flow of ideas \cite{niculae16constructive,zhang16flow} may provide deeper insight into the effect of context on trolling behavior.

Different trolling strategies may also vary in prevalence and severity (e.g., undirected swearing vs. targeted harassment and bullying).
Understanding the effects of specific types of trolling may also allow us to design measures better targeted to the specific behaviors that may be more pertinent to deal with.
The presence of social cues may also mediate the effect of these factors: while many online communities allow their users to use pseudonyms, reducing anonymity (e.g., through the addition of voice communication \cite{davis2002decreasing} or real name policies \cite{cho2013more}) can reduce bad behavior such as swearing, but may also reduce the overall likelihood of participation \cite{cho2013more}.
Finally, differentiating the impact of a troll post and the intent of its author (e.g., did its writer intend to hurt others, or were they just expressing a different viewpoint? \cite{lee2015people}) may help separate undesirable individuals from those who just need help communicating their ideas appropriately.

Future work could also distinguish different types of users who end up trolling.
Prior work that studied users banned from communities found two distinct groups -- users whose posts were consistently deleted by moderators, and those whose posts only started to get deleted just before they were banned \cite{cheng2015antisocial}.
Our findings suggest that the former type of trolling may have been innate (i.e., the user was constantly trolling), while the latter type of trolling may have been situational (i.e., the user was involved in a heated discussion).

\section{Conclusion}
\label{sec:conclusion}
Trolling stems from both innate and situational factors -- where prior work has discussed the former,
this work focuses on the latter, and reveals that both mood and discussion context affect trolling behavior.
This suggests the importance of different design affordances to manage either type of trolling.
Rather than banning all users who troll and violate community norms, also considering measures that mitigate the situational factors that lead to trolling may better reflect the reality of how trolling occurs.

\section{Acknowledgments}
We would like to thank Chloe Kliman-Silver, Disqus for the data used in our observational study, and our reviewers for their helpful comments.
This work was supported in part by a Microsoft Research PhD Fellowship, a Google Research Faculty Award, NSF Grant IIS-1149837, ARO MURI, DARPA NGS2, SDSI, Boeing, Lightspeed, SAP, and Volkswagen.

\balance{}
\bibliographystyle{SIGCHI-Reference-Format}

\end{document}